%% file: article.tex
\documentclass{sig-alternate}

\usepackage[utf8]{inputenc}
\usepackage[english]{babel}

\usepackage{amsmath}
\usepackage{amssymb}
\usepackage{bm}
\usepackage{listings}
\usepackage{color}
\usepackage{url}
\usepackage{subfigure}
\usepackage{multirow}
\usepackage{bussproofs}
\usepackage{algorithmic}
\usepackage{mathtools}
\usepackage{intmacros}
\usepackage{wrapfig}

\usepackage{article}

\usepackage[
  pass,
]{geometry}

\newdef{definition}{Definition}
\newdef{example}{Example}
\newdef{remark}{Remark}

\begin{document}

%
\conferenceinfo{X10'15}{June 14, 2015, Portland, OR, USA}
\CopyrightYear{2015}
\crdata{978-1-4503-3586-7/15/06}

\title{Scalable Parallel Numerical Constraint Solver\\ Using Global Load Balancing}

\numberofauthors{3}

\author{
\alignauthor
Daisuke Ishii \quad \\
       \affaddr{Tokyo Institute of Technology}\\
	   \email{dsksh@acm.org}
\alignauthor
Kazuki Yoshizoe\\
	   \affaddr{The University of Tokyo}\\
	   \email{yoshizoe@acm.org}
\alignauthor
Toyotaro Suzumura\\
       \affaddr{IBM Watson Research Center,}\\
       \affaddr{University College Dublin, JST}\\
       \email{suzumura@acm.org}
}

\maketitle
\begin{abstract}
We present a scalable parallel solver for numerical constraint satisfaction problems (NCSPs).
Our parallelization scheme consists of homogeneous worker solvers, each of which runs on an available core and communicates with others via the global load balancing (GLB) method.
%
The parallel solver is implemented with X10 that provides an implementation of GLB as a library.
In experiments, several NCSPs from the literature were solved and attained up to 516-fold speedup using 600 cores of the TSUBAME2.5 supercomputer.
\end{abstract}

\category{D.1.3}{Concurrent Programming}{Parallel Programming}

\terms{Algorithms, Artificial Intelligence}

\keywords{Interval Analysis, Constraint Programming, Parallel Programming, Global Load Balancing, X10} 

\section{Introduction}

Numerical constraint satisfaction problems (NCSPs, Section~\ref{s:ncsp}) and their dedicated solvers have been successfully applied to problems in the domain of real numbers~\cite{VanHentenryck1997,Ishii2012,CCGIJ2014}.
Given a NCSP with a search space represented by a \emph{box} (i.e., interval vectors), the \emph{branch and prune algorithm} bisects the box or filters an inconsistent portion of the box repeatedly, until finally obtaining a \emph{paving} (i.e., a set of boxes that precisely enclose the solutions set). 
However, the exponential computational complexity of NCSPs limits the number of tractable instances;
therefore, parallelization of NCSP solvers that can scale on a number of cores is a promising approach for the further development of numerical constraint programming~\cite{IYS2014}.

\emph{Search-space splitting} is a simple and efficient approach for parallelization of CSP solvers, yet state-of-the-art implementations are still limited to scaling up to a few hundred cores (Section~\ref{s:survey}).
Recently, Saraswat et al.~\cite{Saraswat2011} have proposed a \emph{global load balancing} framework: a scalable scheme for the global workload distribution and termination detection of irregular parallel computation, which typically applies to the CSP solving process (Section~\ref{s:glb}).
That framework is implemented with X10 and is available in the official distribution of X10 as the GLB library~\cite{Zhang2014}.

In this work, we propose a parallel NCSP solver that uses the GLB library (Section~\ref{s:impl}).
The solver is simply implemented with X10 by adopting the (sequential) constraint propagator of the Realpaver solver as a unit process of GLB.
Section~\ref{s:exp} reports the experimental results obtained when our method was deployed on 600 cores of the TSUBAME2.5 supercomputer.
Optimal configurations of the GLB parameters are analyzed on the basis of those experimental results.

\section{Related Work}
\label{s:survey}

A survey by Gent et al.~\cite{Gent2011} describes existing parallel CSP solvers by classifying them into three categories: 
search-space splitting methods,
cooperative methods for heterogeneous workers (e.g. portfolios and parallel local search)\cite{BH2006}, and
parallelization of the constraint propagation process~\cite{Goldsztejn2010}.
Our work focuses on the first approach.

The main difficulty of the search-space splitting approach lies in the balanced distribution of the sub-trees, which keeps worker solvers active.
When the search tree becomes highly unbalanced, it becomes difficult to predict the appropriate splitting of the search tree, so a dynamic load balancing scheme becomes necessary.
A \emph{work stealing} scheme is typically used for this purpose:
when a worker is starving, it sends a request to other workers, and workloads are communicated in response.
Most existing works experiment with a limited number (e.g., 40\cite{Schulte2000}, 64\cite{Jaffar2004,BH2006}, or 256
\cite{IYS2014,Bergman2014}) 
of processors, and the load balancing tends to assume a central master process, which may limit scalability~\cite{Xie2010,Bergman2014}.
%

A substantial amount of work exists regarding the parallelization of the \emph{branch and bound} algorithm
with both search-space splitting and work stealing~\cite{Grama2003,Luling1996,Otten2010}.
%
Although the branch and bound algorithm resembles the solving process of NCSPs, existing works consider the discrete domain;
our work explores an efficient parallel method that handles the continuous domain with interval computation.

There exist parallel CSP/SAT solvers implemented with X10~\cite{Bloom2012,Bergman2014,IYS2014,Munera2014}. None has yet utilized the GLB library in its implementation.

%
%

\section{Numerical Constraint\\ Programming}
\label{s:ncsp}

Numerical constraint programming is an extension of discrete constraint programming \cite{RBW2006-CP} and uses techniques that are inherited from interval analysis~\cite{Moore1966}. 
Their variable domains are continuous subsets of $\RealSet$: (machine-representable) \emph{intervals} $[a,b] := \{r\in\RealSet ~|~ a \leq r\leq b\}$, where $a$ and $b$ are floating-point numbers, and \emph{boxes} (or vectors of intervals) $([a_1,b_1],\ldots,[a_n,b_n])$.
In the following, boldface characters (e.g., $\vv$) denote intervals and boxes.
$\IntSet$ denotes the set of intervals and $\IntSet^n$ denotes the set of $n$ dimensional boxes.
Numerical constraint solving resorts to \emph{validated} interval computation and the branch and prune scheme:
the solvers evaluate an \emph{interval extension} of a real function in a reliable manner, e.g., an interval extension of addition is computed as $[a_1,b_1]+[a_2,b_2] := [\lfloor a_1+a_2 \rfloor, \lceil b_1+b_2 \rceil]$ where $\lfloor\cdot\rfloor$ and $\lceil\cdot\rceil$ denote the downward and upward rounding mode control,
and the branch and prune algorithm enumerates interval assignments based on the dichotomy principle.

A \emph{numerical constraint satisfaction problem} (NCSP) is defined as a triple $(v,\vv_0,c)$ that consists of
a vector of \emph{variables} $v = (v_1,\ldots, v_{n})$,
an \emph{initial domain} in the form of a box $\vv_0 \in \IntSet^{n}$, and
a \emph{constraint} $c(v) := f(v)=0 \land g(v)<0$, where $f:\RealSet^{n}\to\RealSet^{n_f}$ and $g:\RealSet^{n}\to\RealSet^{n_g}$, i.e., a conjunction of $n_f$ equations and $n_g$ inequalities.
A \emph{solution} of a NCSP is an assignment of its variables $\tilde{v} \in \vv_0$ that satisfies the constraint $c(\tilde{v})$. The \emph{solution set} $\Sigma$ is the region 
$\{\tilde{v}\in \vv_0 ~|~ c(\tilde{v})\}$.
A NCSP is 
\emph{well-constrained} when $n = n_f$,
\emph{under-constrained} when $n > n_f$, and
\emph{over-constrained} when $n < n_f$.
In general, a well-constrained NCSP has a discrete solution set and an under-constrained NCSP has a continuous solution set.

\begin{example} \label{ex}
	We can model the intersection of two disks in the $(v_1,v_2)$ plane as an under-constrained NCSP, where $v := (v_1,\ldots,v_4)$, $\vv_0 = ([-1,1],[-1,1],[0,1],[0,1])$, and
	\[
		c \,\equiv\, (v_1^2 + v_2^2 - v_3,\ (v_1-1)^2 + v_2^2 - v_4) = 0.
	\]
	The solution set projected onto the $(v_1,v_2)$ plane is depicted in Figure~\ref{f:paving}.
\end{example}

\subsection{Branch and Prune Algorithm}
\label{s:bap}

The \emph{branch and prune algorithm}~\cite{VanHentenryck1997} is the standard solving method for NCSPs. 
It takes a NCSP and a precision $\epsilon$ as input and outputs a set of boxes (or \emph{paving}) $\BoxSet$ that approximates the solution set with precision $\epsilon$. 

Figure~\ref{a:bap} presents the algorithm.
%
%
In the main loop at Lines 2--11,
the algorithm first takes the first element of the queue $L$ 
of boxes
and applies the $\Prune$ procedure that shaves boundary portions of the considered box (Line~3).
In this work, we use a basic implementation $\AlgName{HC4Revise}$~\cite{Benhamou1999hullbox} for well-constrained problems;
for under-constrained problems, we use an implementation proposed in \cite{Ishii2012} that provides a verification process based on an interval Newton method combined with $\AlgName{HC4Revise}$.
As a result of $\Prune$, a box becomes either empty, precise enough (its width is smaller than $\epsilon$), verified as an \emph{inner} box of the solution set $\Sigma$, or undecided.
Precise and inner boxes are appended to $S$ (Line~6) and undecided boxes are passed to $\Branch$.
Next, the $\Branch$ procedure bisects the box at the midpoint along a component corresponding to a variable and the sub-boxes are put back into the queue (Line~8).
In this work, we assume that $\Branch$ selects variables in an order that makes the search behave in a breadth-first manner.

\begin{figure}[tb]
\linespread{1}
  \begin{algorithmic}[1]
   \REQUIRE NCSP $\Struct{\VV, \VVInit, \Constr}$, precision $\epsilon$
   \ENSURE set of boxes $\BoxSet$

  \STATE $L \Asn \{\VVInit\}; \BoxSet \Asn \emptyset$
  \WHILE{$L \neq \emptyset$}
    \STATE $\vv \Asn \Prune_\Constr(\Extract(L))$
    \IF{$\vv \neq \emptyset$}
      \IF{$\Wid{\vv} \leq \epsilon \lor \vv$ is an inner box}
        \STATE $\BoxSet \Asn \BoxSet \cup \{\vv\}$
      \ELSE
        \STATE $L \Asn \Insert(L, \Branch(\vv))$
      \ENDIF
    \ENDIF
  \ENDWHILE
  \RETURN{$\BoxSet$}
  \end{algorithmic}
  \caption{Branch and prune algorithm}
  \label{a:bap}
\end{figure}


Realpaver\cite{Granvilliers2006} has been developed as a (sequential) implementation of a NCSP solver.

\begin{figure}
\centering
\includegraphics[width=.24\textwidth]{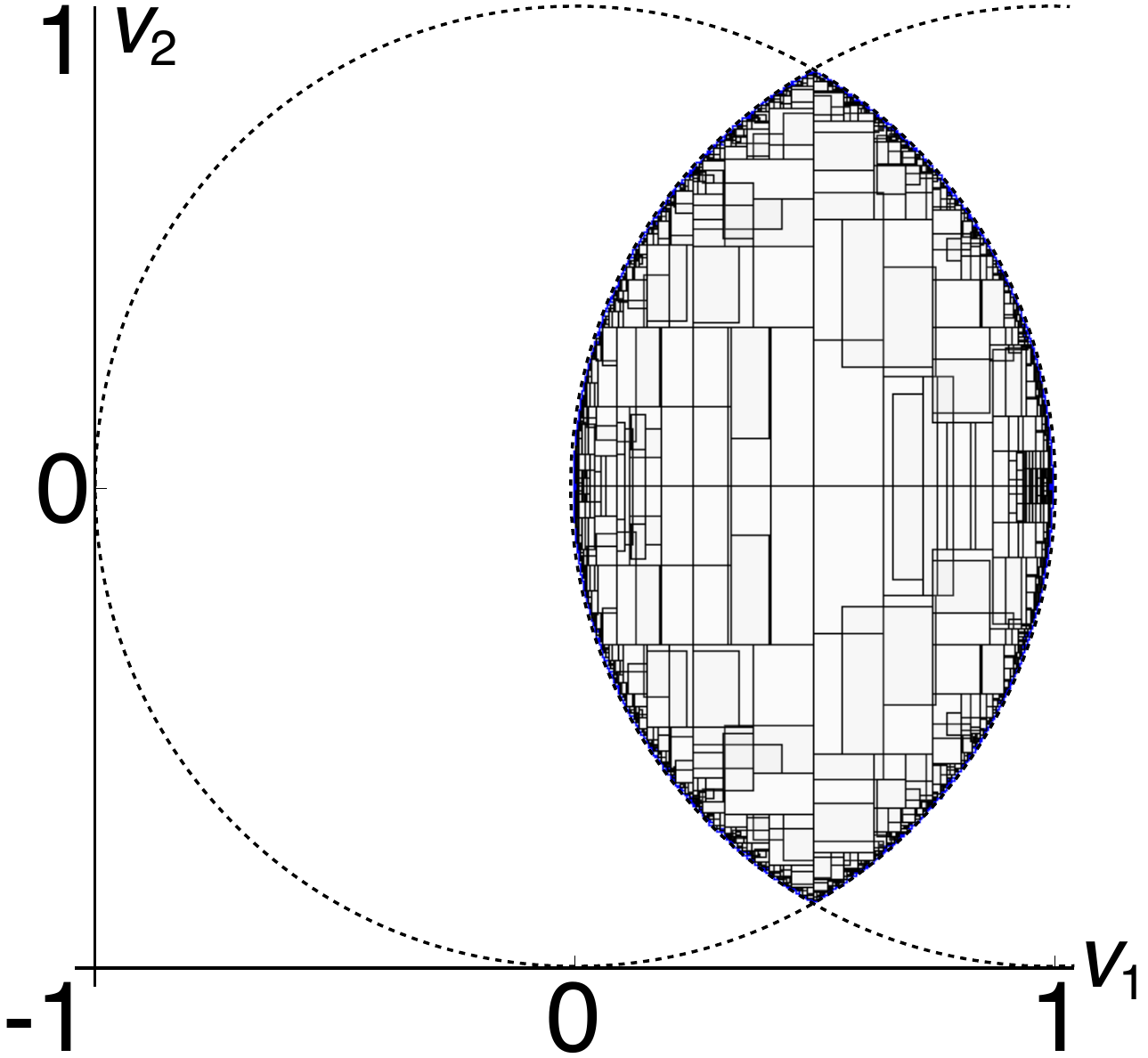} 
\caption{\label{f:paving} Paving of a solution set}
\end{figure}

%
\begin{example}
	A set of boxes enclosing the solution set of the NCSP from Example~\ref{ex}, which is computed with $\epsilon=0.01$, is shown in Figure~\ref{f:paving}.
\end{example}

There are some characteristics that make the parallel solving of NCSPs different from that of other CSPs.
First, computing $\Prune$ is expensive and causes a bottleneck during the solving process; in our experiments, a $\Prune$ call takes around 1ms in average.
Second, $\Prune$ applications result in unbalanced search trees.
Under certain conditions, $\Prune$ contracts a large portion of a considered box (cf. quadratic convergence of the interval Newton methods)
and may even filter out the entire box if the box in question is verified as an inner or totally inconsistent region.
Thus, it is crucial for efficient NCSP solving to execute $\Prune$ at each step while traversing the search tree, which, on the other hand, makes it more difficult to distribute a search path among processors.
Third, the number of solutions is large when a problem is under-constrained and a small $\epsilon$ is given; this causes the search tree to spread toward the bottom.
Last, the search processes for different branches do not require communication; thus, it is safe to run processes on different cores in parallel, without any modification.
%

\section{Parallelization Using\\ Global Load Balancing}
\label{s:parallel}

We parallelize the branch and prune algorithm by splitting and distributing a search space (represented as queue $L$ content) among the workers that run the branch and prune processes homogeneously.
A balanced distribution is not straightforward;
a naive method is to create a frontier of sufficient number of nodes in the search tree, and distribute them evenly across the workers; however, a breadth-first search computation of such a frontier is not efficient because of the time-consuming $\Prune$ process.

The proposed method is implemented simply with X10 and the global load balancing (GLB), which is an efficient scheme for load balancing of irregular tasks.
%

\subsection{X10 GLB Library}
\label{s:glb}

\emph{GLB} is a \emph{global load balancing} library \cite{Zhang2014} in the X10 standard library that implements the lifeline graph work-stealing algorithm \cite{Saraswat2011}.
GLB is suitable for parallelizing irregular tasks, where the workload for each subtask is not predictable, such as search algorithms for AI applications.

GLB computation is performed by multiple cooperative \emph{workers}.
Each worker runs on an X10 place and homogeneously processes a divided workload.
The load balancing between workers is done in two phases:
first, work stealing via requests sent from one worker to randomly selected other workers;
then, work stealing and termination detection via a hyper-cube network of workers called a \emph{lifeline}.
GLB is simply implemented with X10 with configurable parameters and scales up to 16K places when applied to several benchmarks.
For each GLB application, a sequential computation that processes a workload is implemented as an X10 class $\AlgName{TaskQueue}$ and an instance is given to a worker as input.
There are four parameters:
$\PI \in \PosRealSetE$ specifies the lower bound on the time (in seconds) taken by a unit of sequential process\footnote{We modified GLB to use the parameter $\PI$ instead of $\PN$, which specified the number of processed unit tasks.};
$\PW \in \NatSet$ specifies the number of attempted workload steals in the first phase;
and $\PL \in \NatSet$ and $\PZ \in \NatSet$ specify the diameter of the lifeline graph and the number of branches of each node, respectively.
$\PW=0$ turns off the random stealing process.
We assume $\PL^\PZ$ is greater than or equals to the number of workers.
A tight lifeline graph is built by setting $\PL=2$; broadcasting is done in two hops.

\begin{figure}[tb]
\linespread{1}
  \begin{algorithmic}[1]
   \REQUIRE environment $E$, $\AlgName{TaskQueue}$ instance $Q$ 
   \ENSURE task result
   \PARAM $\PI \in \PosRealSetE, \PW, \PL, \PZ \in \NatSet$

  \STATE $\Lifelines := \AlgName{InitLifeline}_E(\PL,\PZ)$

  \REPEAT
  \WHILEC{$Q.\AlgName{process}(\PI)$}{active phase}
  \STATE $\AlgName{DistributeToThieves}_E(Q)$
  \ENDWHILE


  \COMMENT{idle phase 1}
  \FOR{($j := 1$;~ $j \leq \PW \LAnd Q.\mathit{empty}$;~ $j$++)}
  \STATE $\AlgName{TryStealFrom}_E(\AlgName{RandomId}())$
  \ENDFOR


  \COMMENT{idle phase 2}
  \FOR{($j := 1$;~ $j \leq \Lifelines.\mathit{length} \LAnd Q.\mathit{empty}$;~ $j$++)}
  \IF{$\neg\Lifelines(j).\mathit{activated}$}
  \STATE $\Lifelines(j).\mathit{activated} := \True$
  \STATE $\AlgName{TryStealFrom}_E(\Lifelines(j))$
  \ENDIF
  \ENDFOR

  \UNTIL{$Q.\mathit{empty}$}
  \RETURN $Q.\AlgName{getResult}()$
  \end{algorithmic}
  \caption{Worker process}
  \label{a:glb}
\end{figure}
  
A pseudo algorithm in Figure~\ref{a:glb} mimics the worker process.
When the $\AlgName{TaskQueue}$ instance $Q$ contains a workload, a worker becomes active and iterates the loop at Lines~3--5.
A call to the $\AlgName{process}(\PI)$ method of $Q$ invokes a unit sequential computation that should take at least $\PI$ seconds.
$\AlgName{DistributeToThieves}$ sends portions of the available workload (i.e. $Q.\AlgName{split}()$) to other idling workers; otherwise, it signals that the worker has no workload available.
When all workload is processed and property $Q.\mathit{empty}$ becomes true, the worker enters the two-stage idle phase:
at Lines 6--8, the worker randomly selects another worker, sends a request, and waits for the victim's $\AlgName{DistributeToThieves}$ to respond;
at Lines 9--14, the worker sends a request following the lifeline graph.
When the steal succeeds, the loot $\tilde{L}$ is merged by executing $Q.\AlgName{merge}(\tilde{L})$.
When $Q$ is still empty, 
the worker process terminates and outputs the task result.

\subsection{Implementation of NCSP Solver with GLB}
\label{s:impl}

We implement $\AlgName{TaskQueue}$ to encapsulate computation of the branch and prune algorithm. $\AlgName{TaskQueue}$ holds the queue $L$ of undecided boxes and the solution set $S$.
Initially, a worker possesses the initial domain $\VVInit$ in $L$ and the queues of other workers are left empty.
Methods of $\AlgName{TaskQueue}$ are implemented as follows:
\begin{itemize}
\setlength{\parskip}{0pt}
\setlength{\itemsep}{0pt}
	\item $\AlgName{process}(\PI)$ computes the main loop of the branch and prune algorithm until the time $\PI$ elapses.
		For $\Prune$, the C++ implementation of Realpaver is used.
	\item $\AlgName{split}()$ divides $L$ equally into two and returns a portion.
	\item $\AlgName{merge}(\tilde{L})$ is given a set $\tilde{L}$ of boxes and appends the boxes to $L$.
	\item $\AlgName{getResult}()$ returns $|S|$. Our implementation does not gather $S$ in one place, thus avoiding unnecessary overhead. Indeed, $S$ might be better distributed in the post-process of many applications.
\end{itemize}

The implementation consists of about 1,000 lines of X10 and 2,400 lines of C++ code.
In NCSP applications, the solving process must be tweaked by trying several combinations of $\Prune$ implementations and search strategies.
In this respect, our simple X10 framework that is interfaced with C++ solver implementations will serve as a practical tool.

\section{Performance Evaluation}
\label{s:exp}

We evaluated our parallel NCSP solver with several problems from existing literature.
The experiments were operated on the TSUBAME2.5 supercomputer,
%
%
which is a supercomputer at Tokyo Institute of Technology.\footnote{\url{http://tsubame.gsic.titech.ac.jp/en}}
Each node of TSUBAME2.5 has two Intel Westmere EP 2.93GHz processors (12 cores in total) and 54GB of local memory.
We used 50 nodes; thus, each experiment was run with up to 600 X10 places on 600 cores.
We used native X10 version~2.4.3.2 and its MPI backend (based on Open MPI~1.6.5).
%

\subsection{Experimental Results}

We solved four instances of two well-constrained (WC) problems taken from \cite{Hentenryck1998,GGJ2013} and six instances of two under-constrained (UC) problems taken from \cite{Ishii2012,CCGIJ2014}.
%
For each of the first three problems, we prepared two instances by varying the number of variables and constraints.
For the UC problems, we solved with two different precisions.
Every instance was solved with the following seven GLB parameter configurations:
\begin{enumerate}
\setlength{\parskip}{0pt}
\setlength{\itemsep}{0pt}
	\item[(1)] $\PI=0.001$s, $\PL=2$, and $\PW=0$. 
	\item[(2)] $\PI=0.001$s, $\PL=2$, and $\PW=1$. 
	\item[(3)] $\PI=0.001$s, $\PL=2$, and $\PW=\PZ$. 
	\item[(4)] $\PI=0.001$s, $\PL=P$, and $\PW=0$. 
	\item[(5)] $\PI=0.001$s, $\PL=P$, and $\PW=\PZ$. 
	\item[(6)] $\PI=0.1$s, $\PL=2$, and $\PW=0$. 
	\item[(7)] $\PI=0.1$s, $\PL=2$, and $\PW=\PZ$. 
\end{enumerate}
$\PZ$ was set as 
$\lceil \log_\PL P \rceil$ (such that $\PL^\PZ \geq P$).
%

\begin{table*}[t]
\centering
  \caption{\label{t:results} Considered problems and experimental results} 
    \begin{tabular}{l|c|r|r|r|c|r|r|r|r|r} \hline \hline
	  problem & size & $\epsilon$ & \# sol & \# br &
	  $\PW$ & $t_1$ & $t_{300}$ & $t_{600}$ &
	  ar${}_{600}$ & \# sb${}_{600}$ \\
	  \hline
	  Economics & 8 & $10^{-8}$ & 8 & 63 478 &
	  0 & 58s & \underline{0.40s} & \underline{0.41s} &
	  64\% & 47 000 \\
	  (\textit{eco8}) & & & & &
	  1 & & 0.78s & 0.77s &
	  25\% & 27 500 \\
	  \hline
	  Economics & 10 & $10^{-8}$ & 16 & 3 614 945 &
	  0 & 5 970s & 22.0s & 11.8s &
	  88\% & 2 550 000 \\
	  (\textit{eco10}) & & & & &
	  1 & & \underline{21.4s} & \underline{11.5s} &
	  93\% & 1 150 000 \\
	  \hline
	  Periodic orbits & 48 & $10^{-8}$ & 2 939 & 28 742 &
	  0 & 1 330s & 8.5s & 5.0s &
	  58\% & 34 800 \\
	  (\textit{henon24}) & & & & &
	  1 & & \underline{6.8s} & \underline{4.4s} &
	  63\% & 19 000 \\
	  \hline
	  Periodic orbits  & 56 & $10^{-8}$ & 16 105 & 174 446 &
	  0 & 12 530s & 60.2s & 31.2s &
	  65\% & 201 000 \\
	  (\textit{henon28}) & & & & &
	  1 & & \underline{45.7s} & \underline{25.1s} &
	  87\% & 81 000 \\
	  \hline

	  2D sphere & 2+2 & $0.004$ & 312 064 & 364 961 &
	  0 & 122s & \underline{0.6s} & \underline{0.5s} &
	  75\% & 295 000 \\
	  and plane & & & & &
	  1 & & 0.8s & 0.9s &
	  39\% & 153 000 \\
	  \cline{3-11}
	  (\textit{sp2-2}) & & $0.001$ & 2 490 988 & 2 936 705 &
	  0 & 780s & \underline{3.8s} & \underline{2.2s} &
	  87\% & 2 300 000 \\
	  & & & & &
	  1 & & 3.9s & 2.6s &
	  78\% & 955 000 \\
	  \hline
	  4D sphere & 2+4 & $0.004$ & 1 459 225 & 2 488 689 &
	  0 & 1 202s & 7.0s & \underline{3.8s} &
	  85\% & 1 790 000 \\
	  and plane & & & & &
	  1 & & \underline{6.6s} & 4.1s &
	  85\% & 662 000 \\
	  \cline{3-11}
	  (\textit{sp2-4}) & & $0.001$ & 11 759 158 & 20 082 197 &
	  0 & 12 800s & 52s & 27s &
	  92\% & 12 850 000 \\
	  & & & & &
	  1 & & \underline{50s} & \underline{26s} &
	  97\% & 4 600 000 \\
	  \hline

	  3-RPR robot & 3+3 & $0.2$ & 1 488 388 & 1 936 939 &
	  0 & 598s & \underline{2.8s} & \underline{1.7s} &
	  80\% & 1 550 000 \\
	  (\textit{3rpr}) & & & & &
	  1 & & 2.9s & 2.1s &
	  71\% & 675 000 \\
	  \cline{3-11}
	  & & $0.1$ & 5 649 780 & 7 186 845 &
	  0 & 2 135s & 9.0s & \underline{5.0s} &
	  86\% & 5 600 000 \\
	  & & & & &
	  1 & & \underline{8.7s} & 5.2s &
	  88\% & 2 300 000 \\
	  \hline
    \end{tabular}
\end{table*}

\begin{figure}[tp]
\centering
  \subfigure[\textit{eco8} (left); \textit{eco10} (right).]
  { \begin{minipage}{.45\textwidth}
    \centering
\includegraphics[height=60pt]{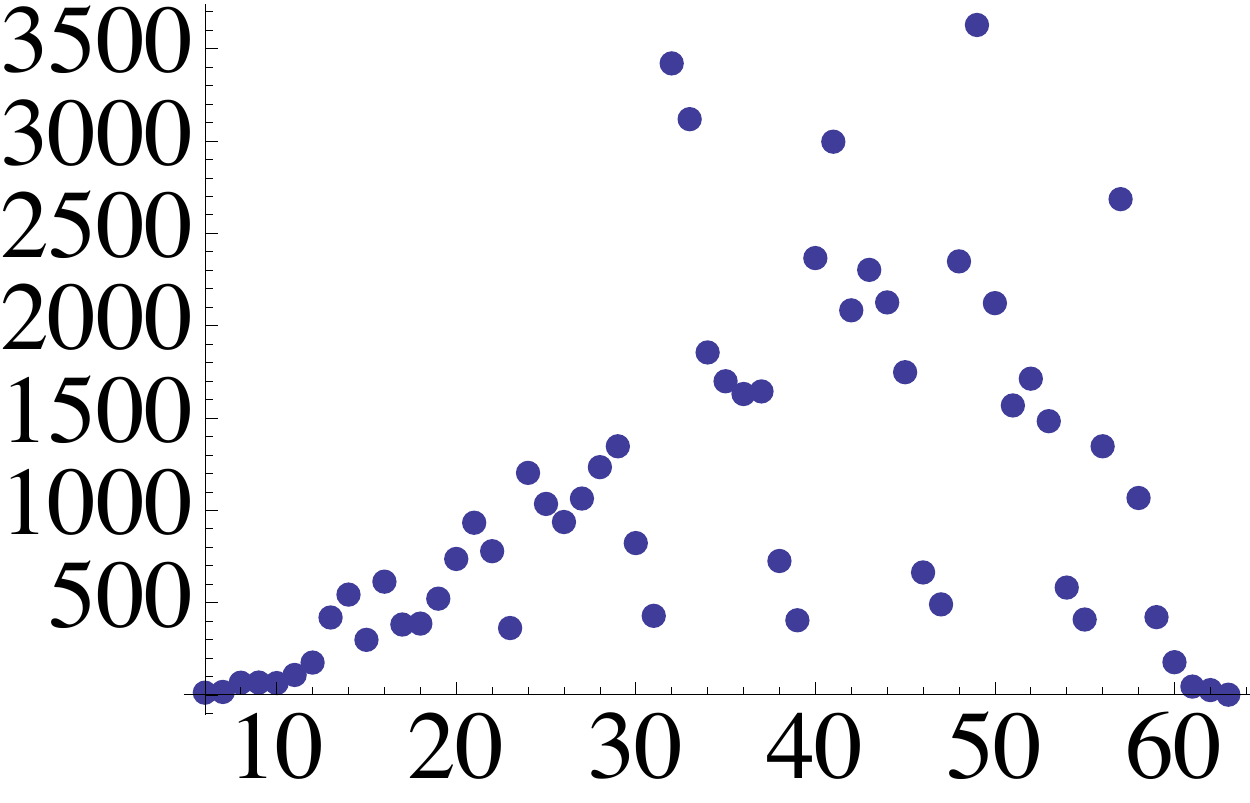}
\hspace{1em}
\includegraphics[height=60pt]{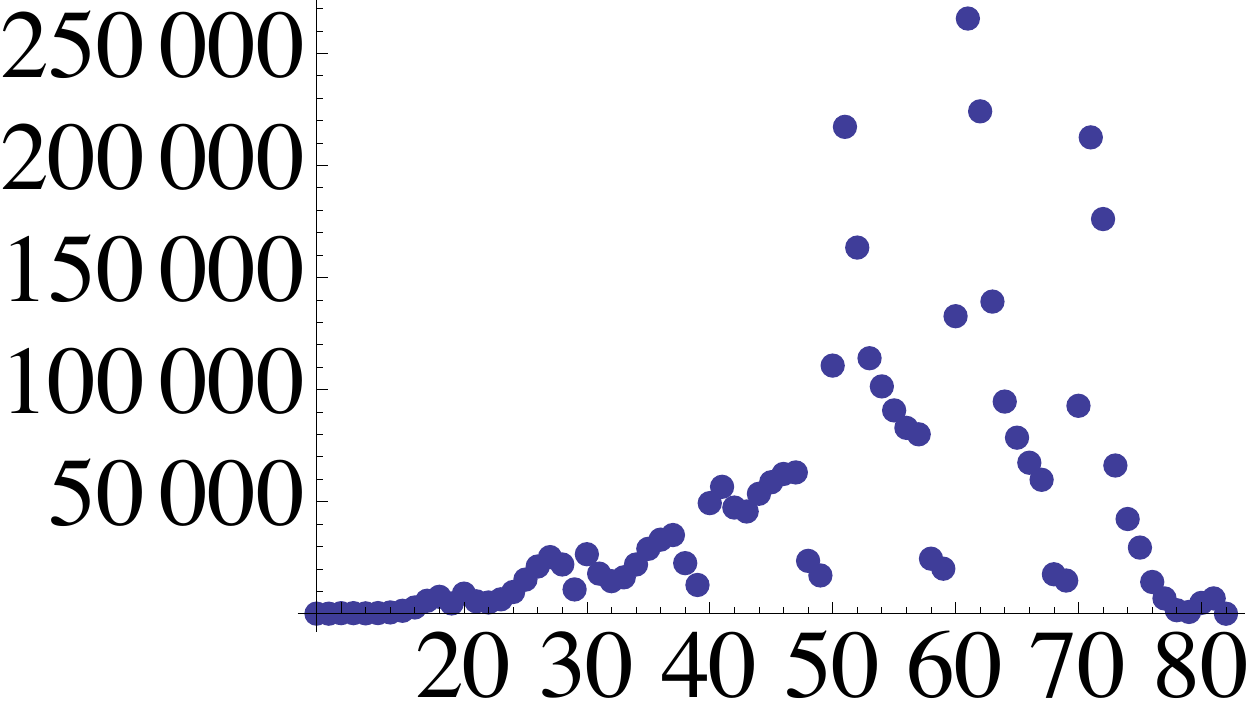}
    \end{minipage}}
  \subfigure[\textit{sp2-2}, $\epsilon=0.004$ (left); \textit{sp2-4}, $\epsilon=0.001$ (right).]
  { \begin{minipage}{.45\textwidth}
    \centering
\includegraphics[height=60pt]{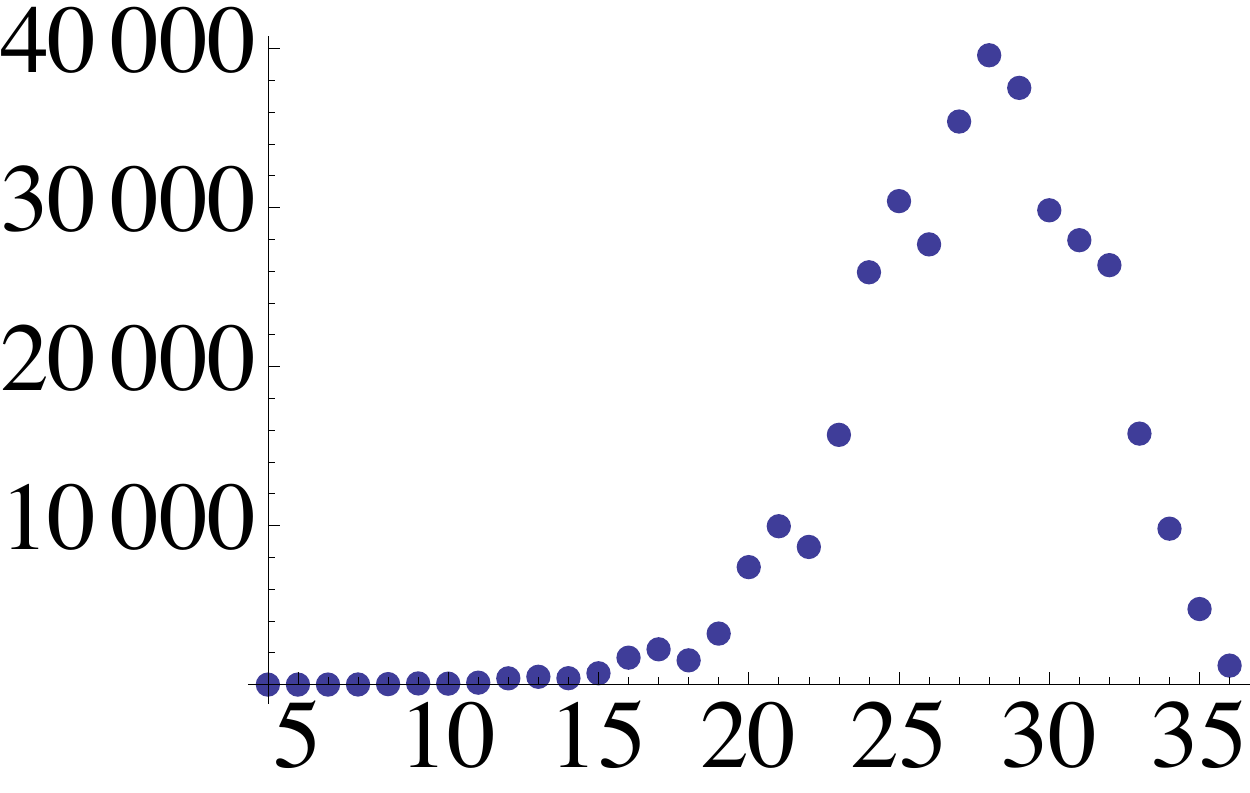}
\hspace{1em}
\includegraphics[height=60pt]{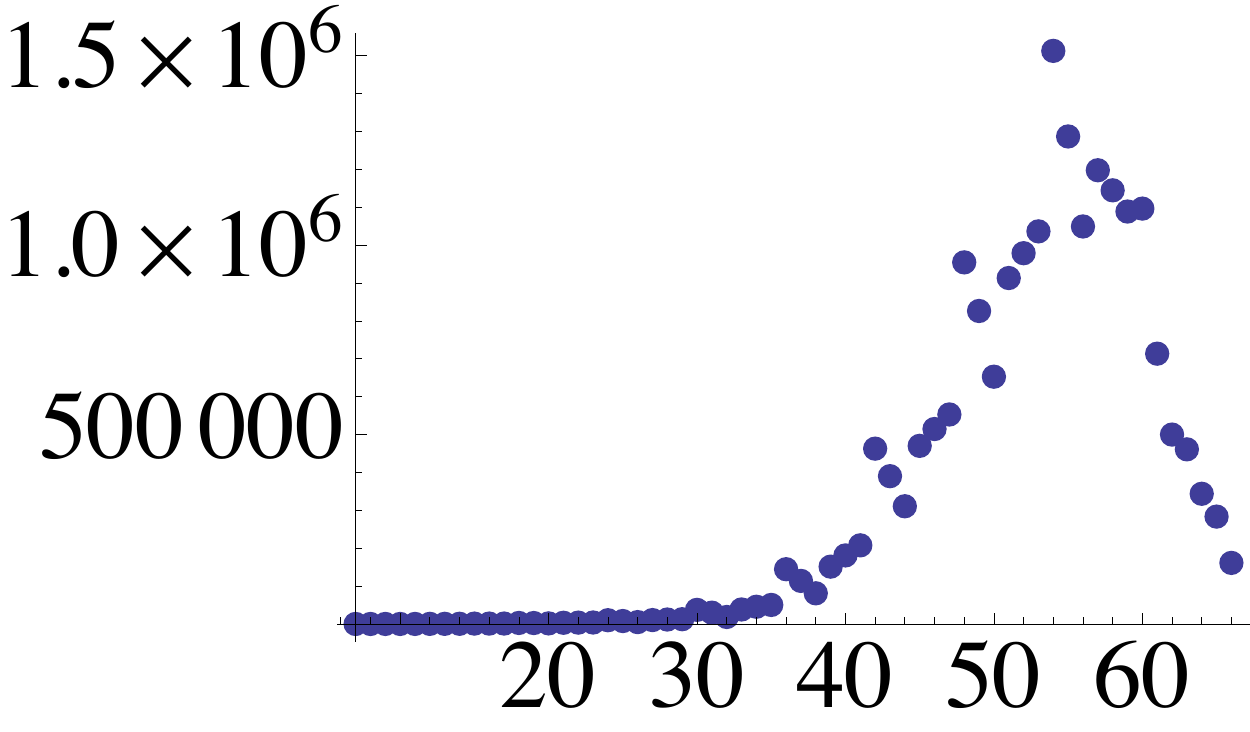}
    \end{minipage}}
	\vspace{-.5em}
	\caption{\label{f:tree} Number of search paths (vertical axis) per depth (horizontal axis)}
\end{figure}

Specification of the instances and experimental results using configurations (1) and (2), which were performed most efficiently, are shown in Table~\ref{t:results}.
The columns ``problem'', ``size'', ``$\epsilon$'', ``\# sol'', ``\# br'', and ``$\PW$'' represent the name of the problem, size (i.e., the number of variables $n$; for UC problems, the number of equality constraints $n_f$ that are separated with `+'), precision, number of solutions, number of branches, and value of $\PW$, respectively.
The rest of the columns provide some experimental results.
$t_j$ represents the running time using $j$ X10 places (best timings are underlined).
$\text{ar}_{600}$ represents the mean of the ratio of active time versus the total solving time at each place when computed with 600 places.
$\text{\# sb}_{600}$ represents the total number of sent boxes from 600 places for load balancing.
Figure~\ref{f:tree} shows the number of paths per depth in each search tree of the four instances.
Figure~\ref{f:su} illustrates the speedups of the parallel solving processes for the seven instances.
Figure~\ref{f:cpu} illustrates the fraction of the three worker states within the CPU timing for the two instances.
Each layer, from bottom to top, corresponds to the time taken for $\Prune$, $\AlgName{DistributeToThieves}$, and the idle phase (Lines~6--14 in Figure~\ref{a:glb}), respectively.\footnote{The timings for $\Prune$ differed per number of places. As a reason, we predicted and confirmed that the CPU cache hit ratio differed in the parallel processes.}

\subsection{Discussion}

\input{graphs.tex}

In the experiments, our parallel solver scaled up to 600 places/cores and achieved up to 516-fold speedup (an efficiency of 0.84).

As can be seen from Figure~\ref{f:tree}, each of the search trees for the instances considered has a number of paths whose lengths are close to the height of the tree; thus, a certain level of parallelism exists.
Furthermore, comparing the graphs of the different instances of the same problem, we see that the shapes of graphs are similar, but the size of the tree of larger instances increases exponentially, indicating that parallelization should be easier for larger instances.

For most instances (excluding the large ones), the best speedups were accomplished with configuration (1): the parallel process with the most frequent load balancing that used the lifeline graph with the broadest distribution and did not perform random stealing.
The efficiency of load balancing is evident in the active ratio of workers (see $\text{ar}_{600}$ in Table~\ref{t:results} and the left-hand figures of Figure~\ref{f:cpu}).
Despite its large communication overhead (see the last column of Table~\ref{t:results}), load balancing that used the lifeline resulted in quick workload distribution and termination.

For large instances, such as \textit{eco10} (Figure~\ref{f:su:eco}, right), \textit{henon28}, and \textit{sp4} ($\epsilon=0.001$; Figure~\ref{f:su:sp}, right), configurations (2), (3), and (5), which had frequent random stealing, outperformed configuration (1).
Their performance also appeared in the active ratio (Figure~\ref{f:cpu:eco}, right).
Among these configurations, configuration (2) performed the best probably because of its quick termination process.

Regarding time interval $\PI$ of the load balancing, the most frequent setting, $\PI=0.001$s, performed well. With this setting, workload was distributed between almost every call to $\Prune$ which takes around 1ms on average and, in total, occupies around 90\% of the running time.

Configuration (4) always performed poorly.
Its load balancing, which used a lifeline formed as a 1D hyper cube (i.e. ring), was slow and led to its poor performance.
Its performance improved dramatically by enabling the random stealing process (configuration (5)).

In some experiments, such as \textit{sp4} ($\epsilon=0.004$; Figure~\ref{f:su:sp}, middle), \textit{3rpr} ($\epsilon=0.1$; Figure~\ref{f:su:3rpr}, right), and \textit{3rrr} ($\epsilon=0.1$), configuration (1) scaled well and outperformed other configurations when using 600 cores.
It would appear that the random stealing process, when using many cores, suffered from large communication overhead;
such large overhead can be confirmed with configuration (2), shown in the right-hand side of Figure~\ref{f:cpu:3rpr}.

Finally, the running times of some experiments were quite short. For example, \textit{eco8} and \textit{sp2-2} ($\epsilon=0.004$) took 58s and 122s, respectively, with a single core, and certain speedups were achieved: 141- and 252-fold with 600 cores.

\section{Conclusions}

In this work, we show that the parallelization of the branch and prune search is a good application of the X10 GLB framework.
In the experiments, we achieved nearly linear speedups up to 600 X10 places/cores and are expected to be able to scale further.
In future work, we plan further experiments on realistic problems including optimization problems, using a greater number of cores. 

\paragraph{Acknowledgments}
\noindent
This work was partially funded by JSPS (KAKENHI\\ 25880008, 15K15968, 25700038, 26280024, and 23240005) and JST ERATO Project.

%
\bibliographystyle{abbrv}
\bibliography{parallel.bib}
%
%

\end{document}

%% file: graphs.tex
\begin{figure*}[p]
\begin{center}
  \includegraphics[width=0.7\textwidth]{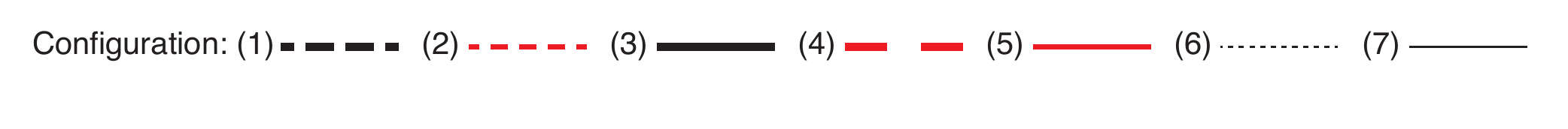}

  \vspace{-2em}
  \subfigure[\label{f:su:eco} \textit{eco8} (left), \textit{eco10} (right)]
  { \begin{minipage}{\textwidth}
    \begin{center}
	  \includegraphics[height=0.22\textwidth]{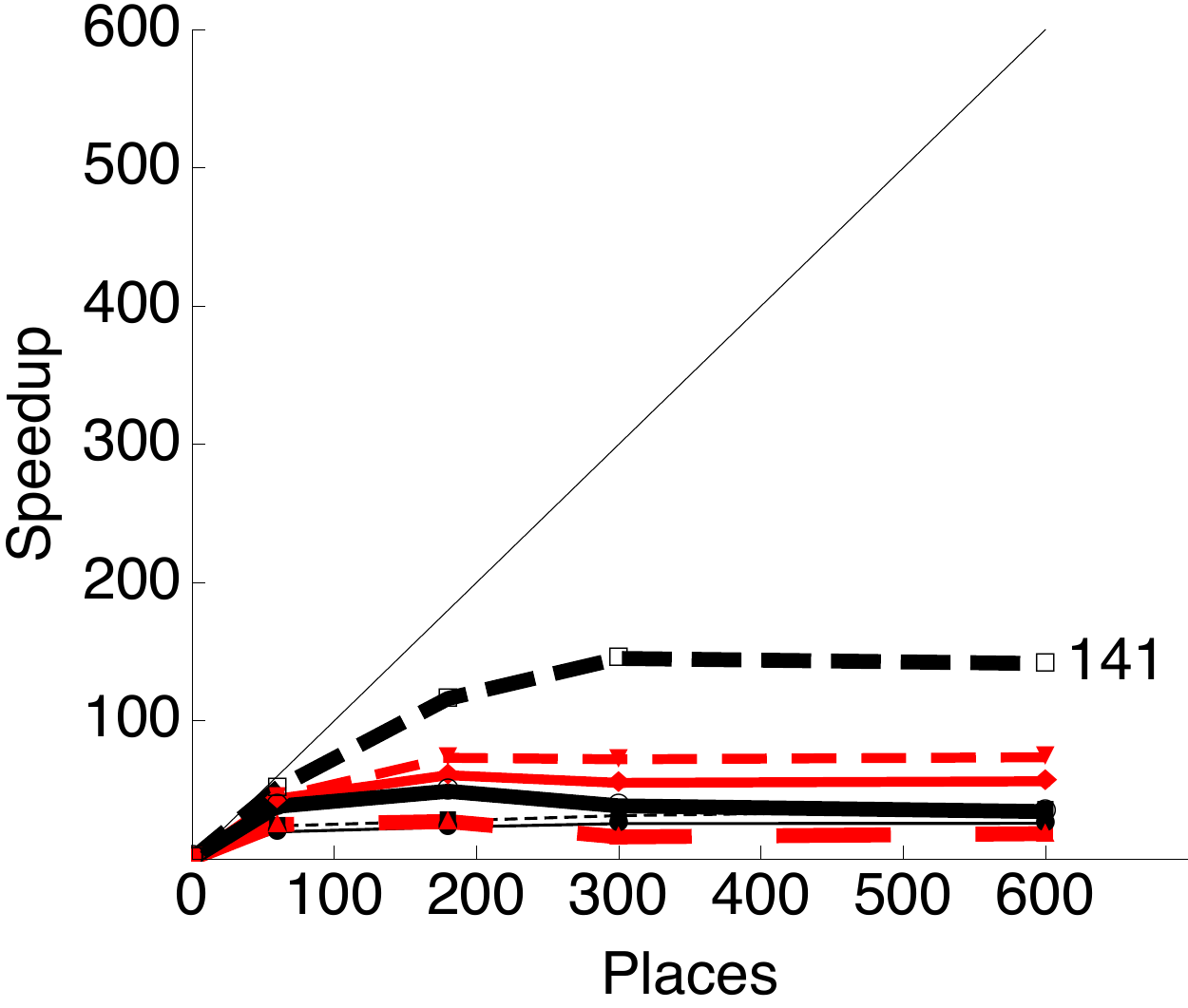}
	  \hspace{1em}
	  \includegraphics[height=0.22\textwidth]{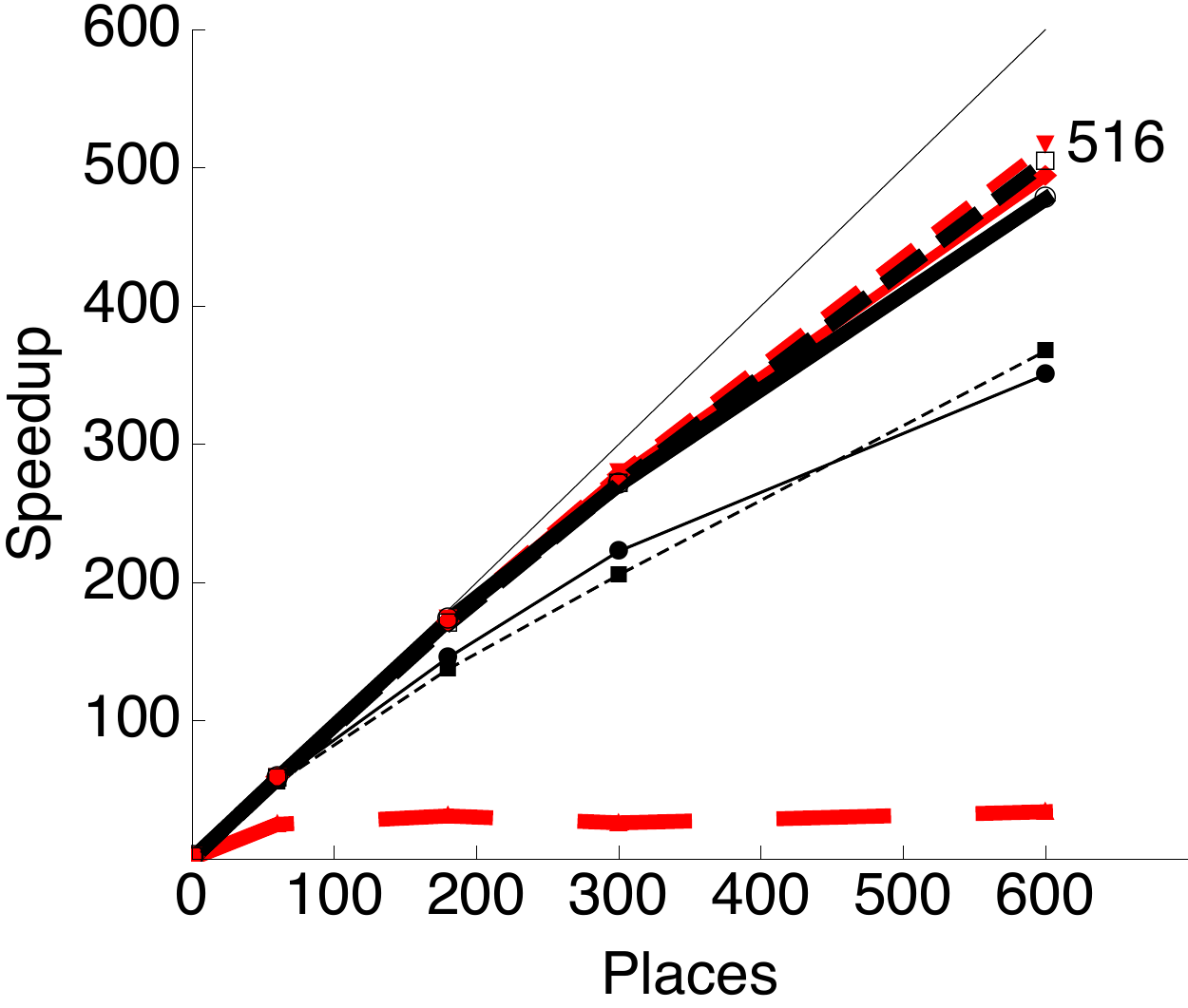}
    \end{center}
    \end{minipage} }
  %
  \subfigure[\label{f:su:sp} \textit{sp2-2}, $\epsilon=0.004$ (left); \textit{sp2-4}, $\epsilon=0.004$ (middle); \textit{sp2-4}, $\epsilon=0.001$ (right)]
  { \begin{minipage}{\textwidth}
    \begin{center}
	  \includegraphics[height=0.22\textwidth]{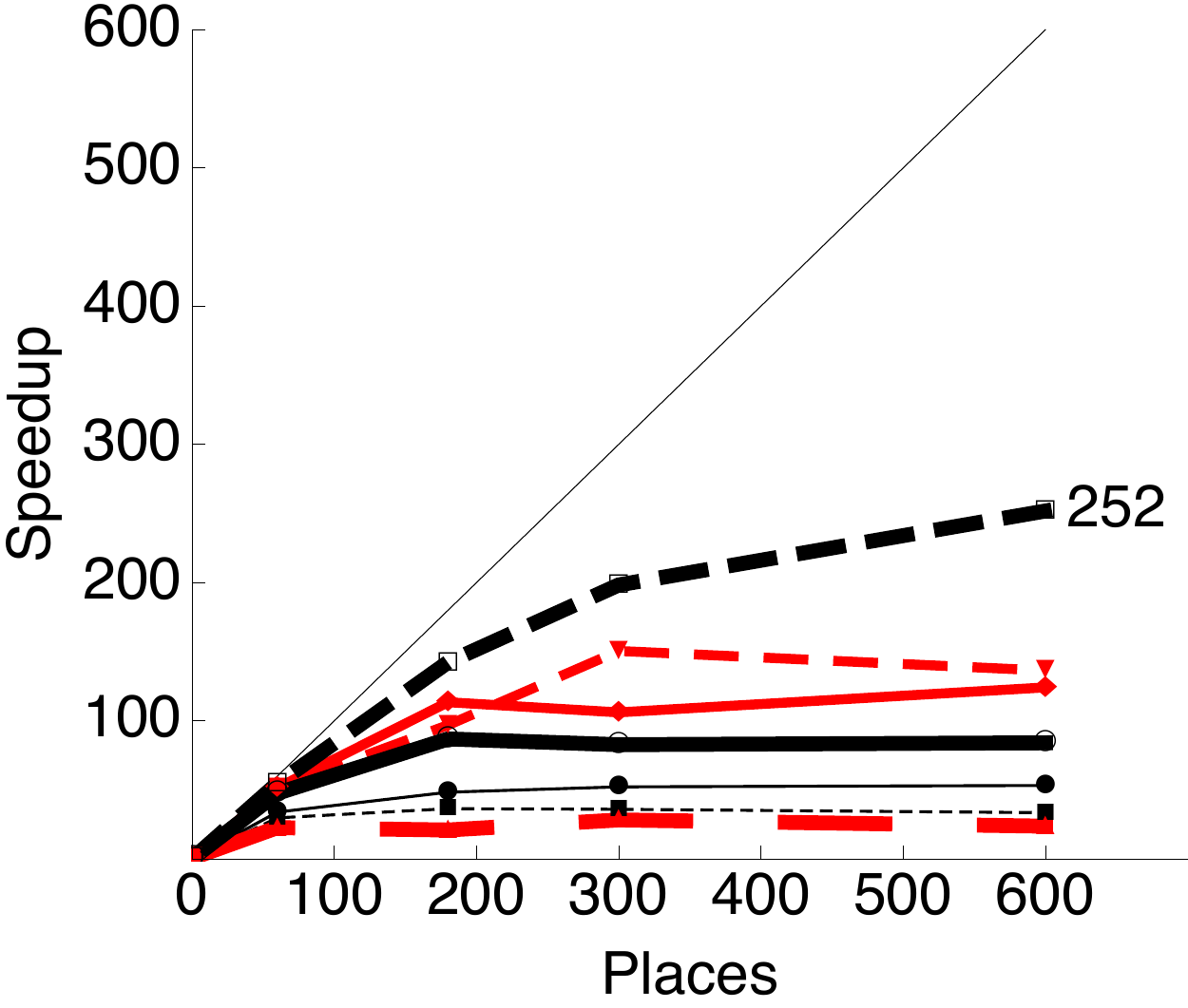}
	  \includegraphics[height=0.22\textwidth]{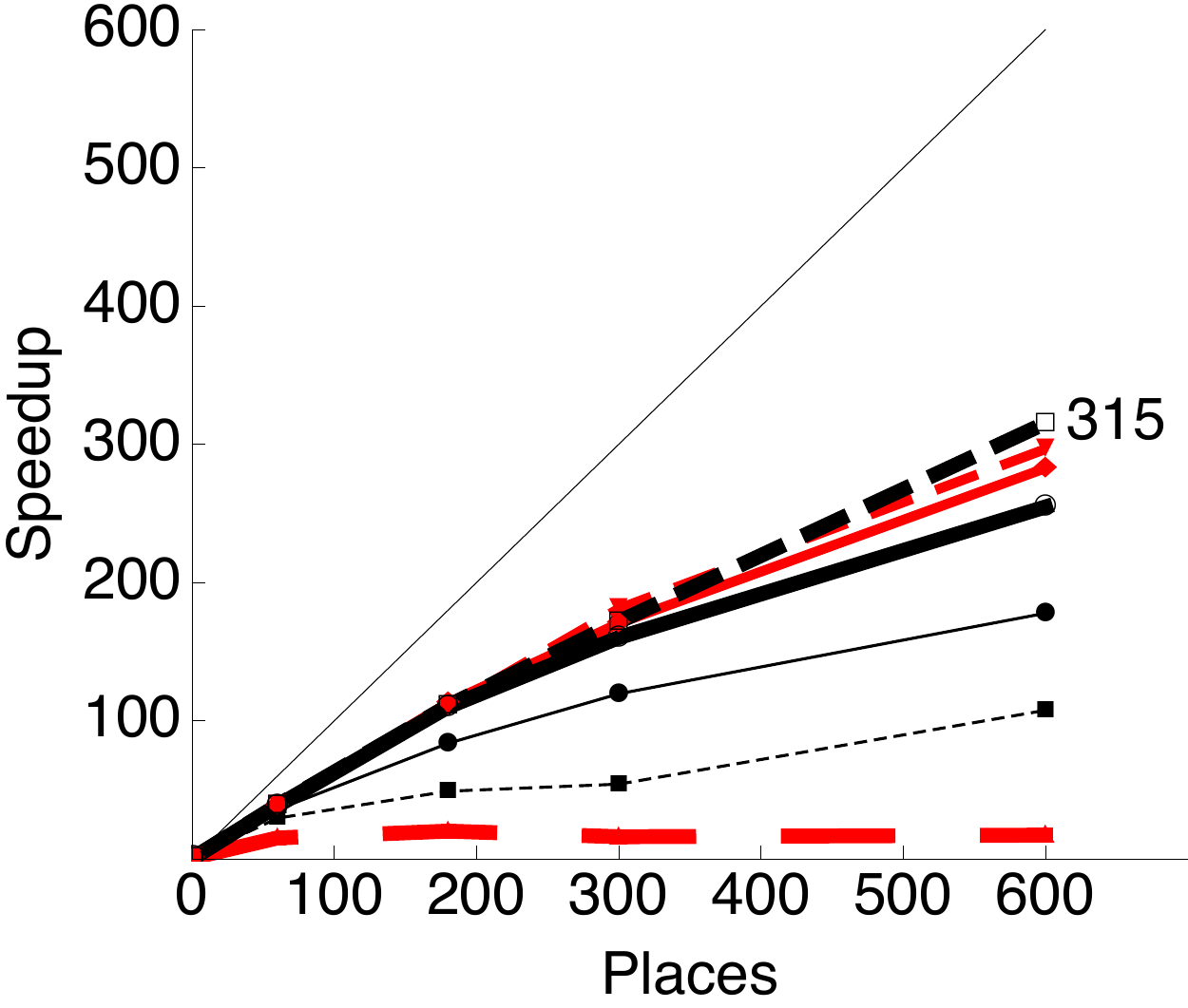}
	  \includegraphics[height=0.22\textwidth]{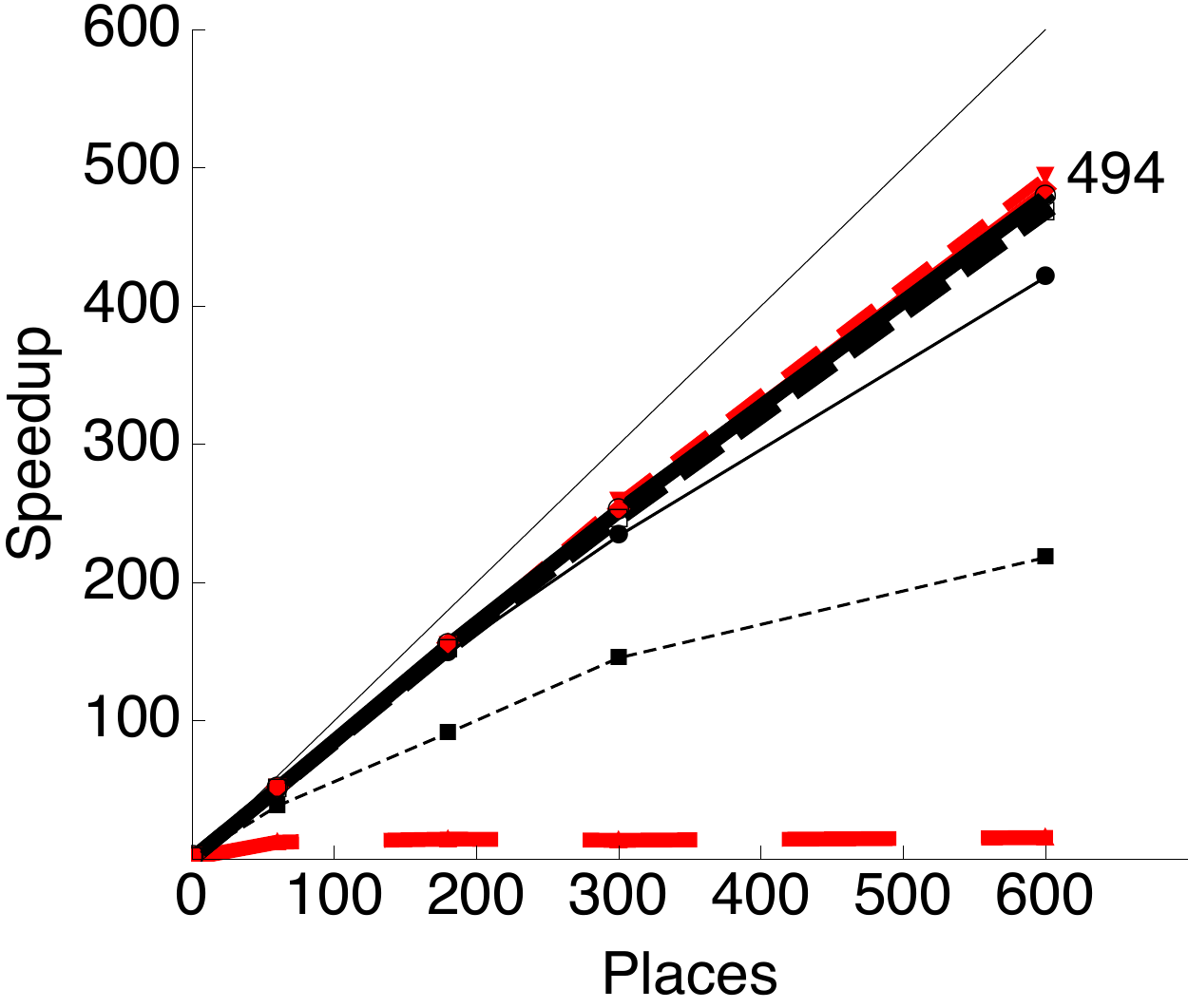}
    \end{center}
    \end{minipage} }
  \subfigure[\label{f:su:3rpr} $\textit{3rpr}, \epsilon=0.2$ (left), $\epsilon=0.1$ (right)]
  { \begin{minipage}{\textwidth}
    \begin{center}
	  \includegraphics[height=0.22\textwidth]{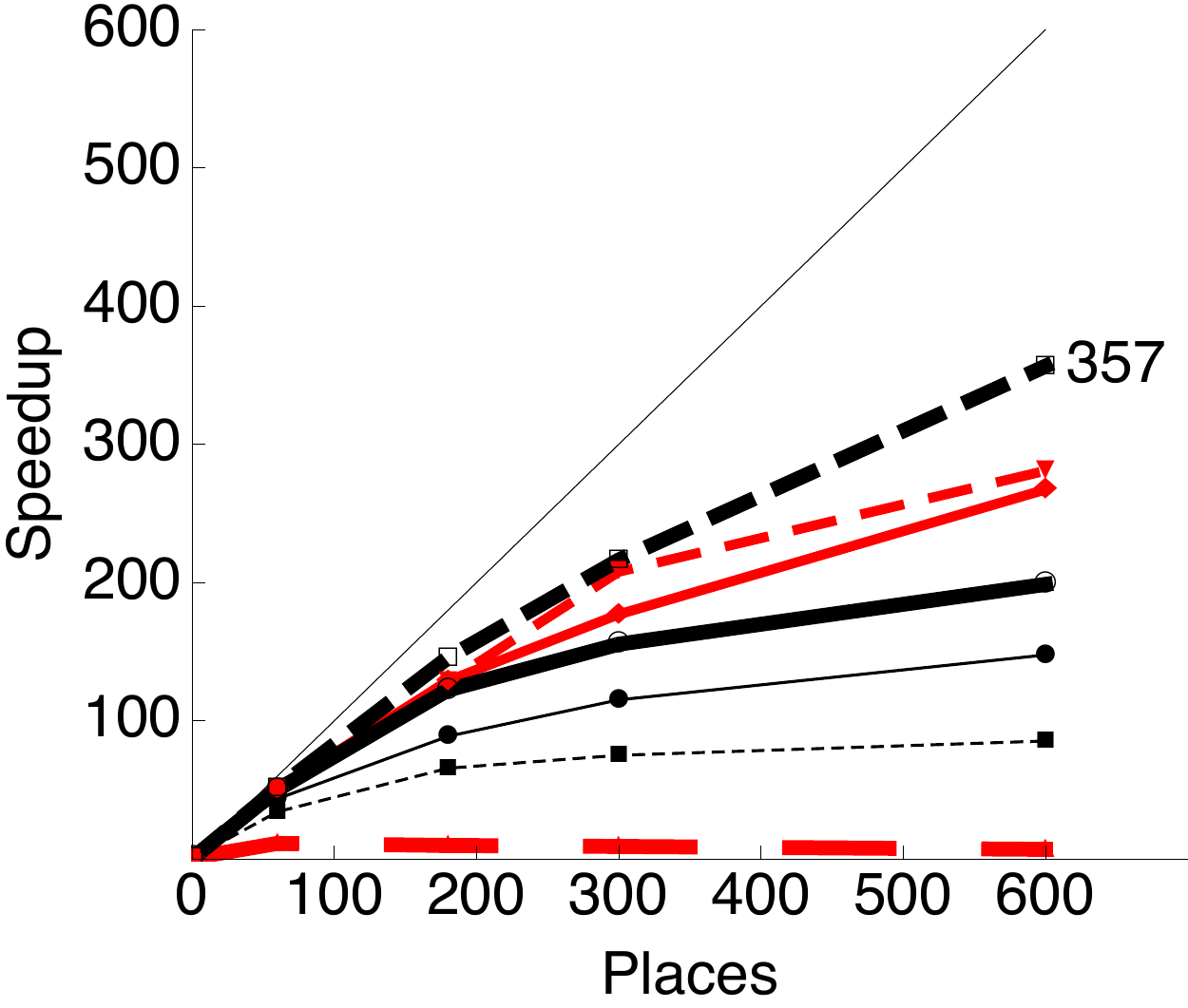}
	  \hspace{1em}
	  \includegraphics[height=0.22\textwidth]{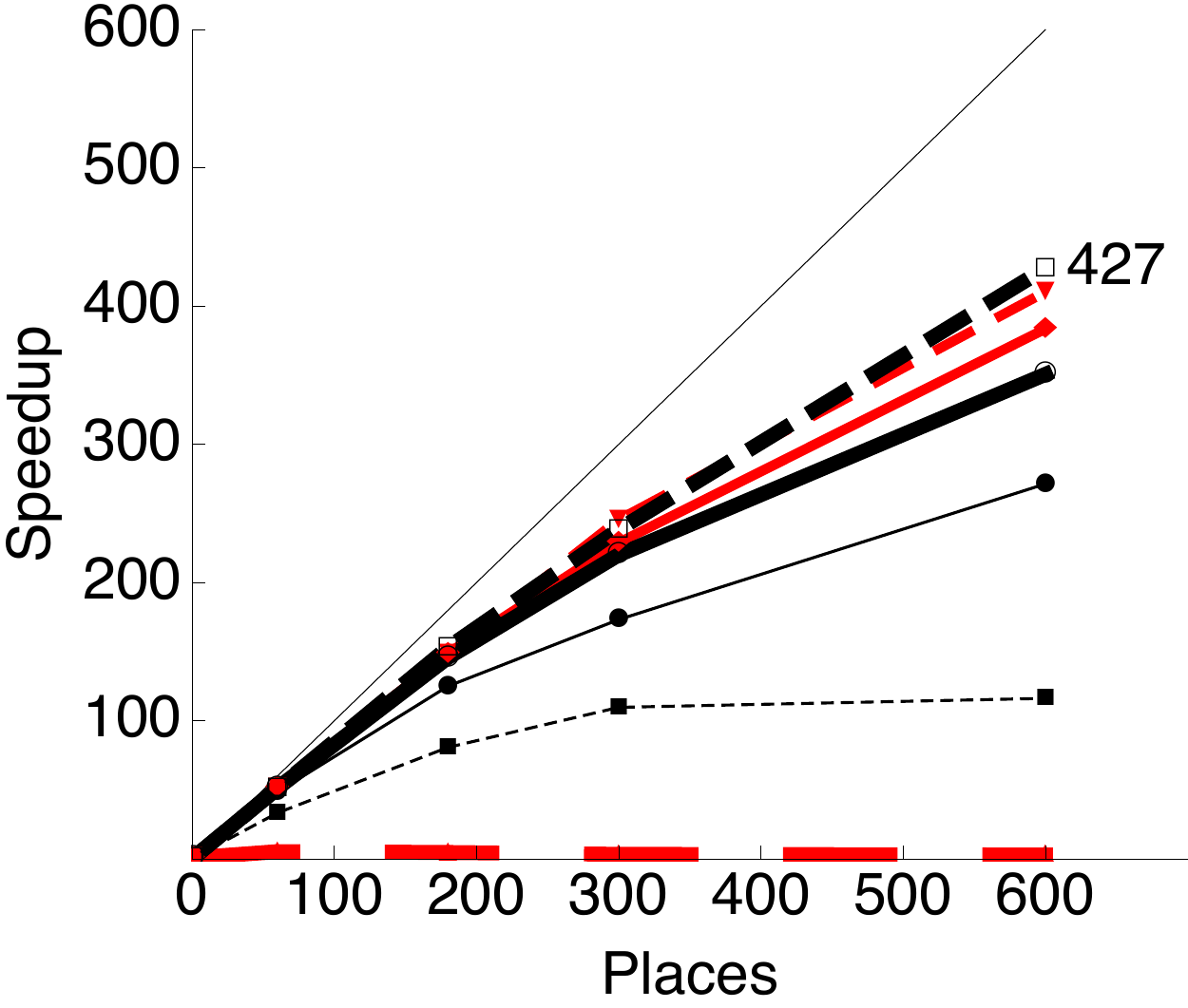}
    \end{center}
    \end{minipage}}
  \vspace{-1.5em}
  \caption{\label{f:su} Speedup of the solving process}
\end{center}
\end{figure*}



\vspace{-1em}

\begin{figure*}[p]
\begin{center}
  \subfigure[\label{f:cpu:eco} \textit{eco8} (left), \textit{eco10} (right)]
  { \begin{minipage}{\textwidth}
    \begin{center}
	  \includegraphics[height=0.1\textheight]{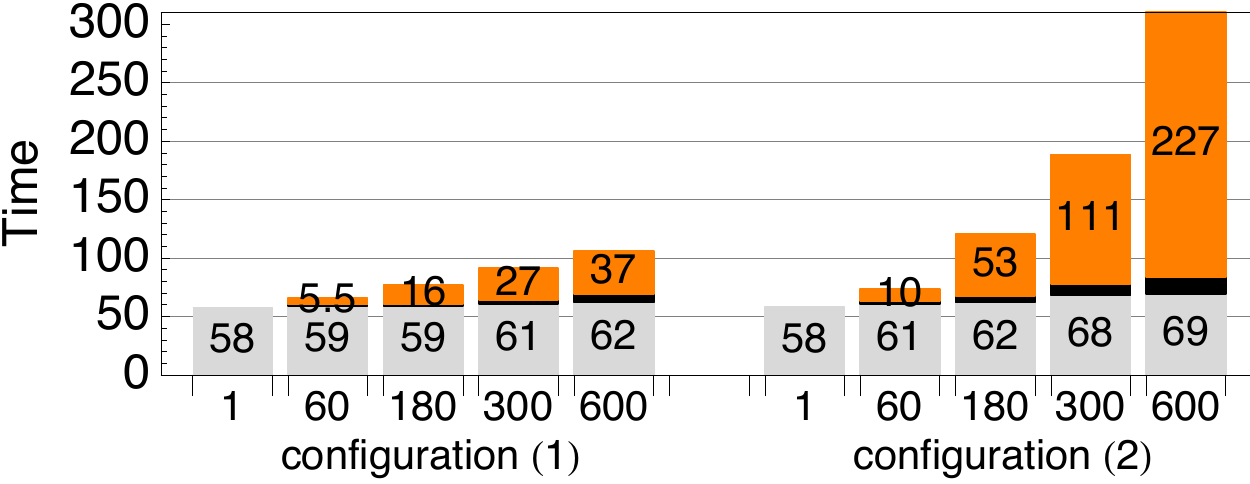}
	  \hspace{1em}
	  \includegraphics[height=0.1\textheight]{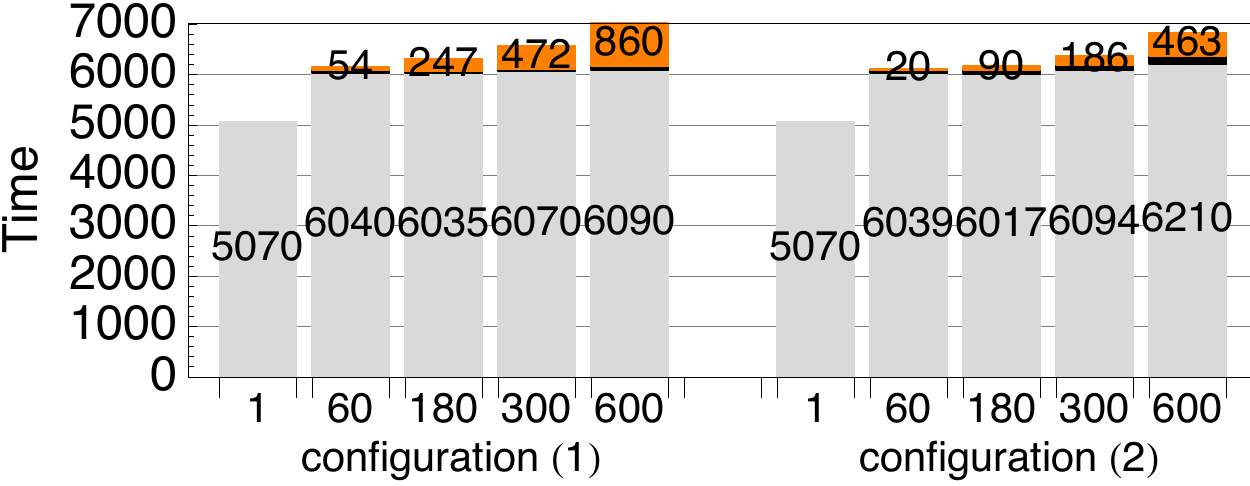}
    \end{center}
    \end{minipage}}
  %
  %
  \subfigure[\label{f:cpu:3rpr} \textit{3rpr}, $\epsilon=0.2$ (left), $\epsilon=0.1$ (right)]
  { \begin{minipage}{\textwidth}
    \begin{center}
	  \includegraphics[height=0.1\textheight]{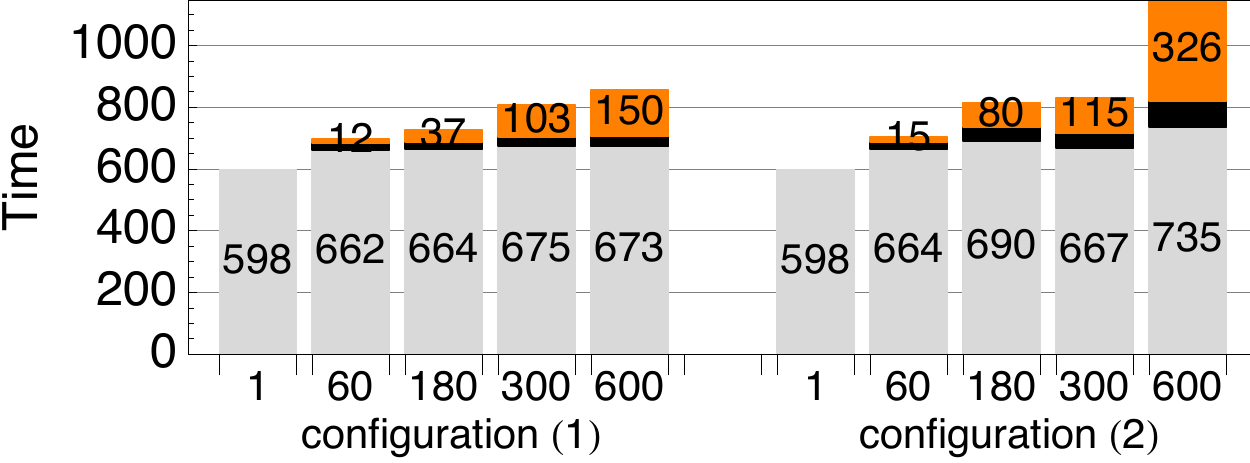}
	  \hspace{1em}
	  \includegraphics[height=0.1\textheight]{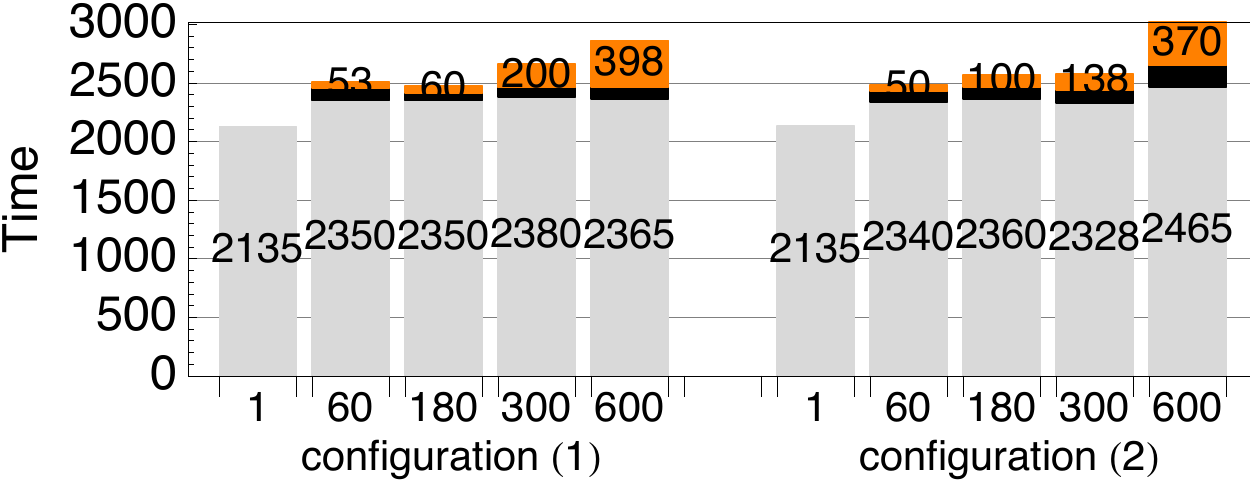}
    \end{center}
    \end{minipage}}
  %
  %
  %
  \vspace{-1.5em}
  \caption{\label{f:cpu} Breakdown of CPU time with timings for pruning, load sending, and idling}
\end{center}
\end{figure*}
